\documentclass[twocolumn,twoside,slac_two]{revtex4}
\usepackage{graphicx}
\usepackage{fancyhdr}
\begin{document}
\setlength{\topmargin}{+0.3in}     
\title{Charm Mixing and CP-violations - Theory}

%

\author{E. Golowich}
\affiliation{Physics Department, University of Massachusetts, Amherst, MA 01003 USA}

\begin{abstract}
This document describes my talk at the $6^{\rm th}$ 
Flavor Physics and CP Violation Conference held at Taiwan National 
University (5/5/08-5/9/08).  I begin by commenting on the most recent 
experimental compilation of $D^0$ mixing data, emphasizing 
the so-called `strong phase' issue.  This is followed by a review 
of the theory underlying charm mixing, both Standard Model and New 
Physics.  The mechanism of R$_{\rm P}$-violation is used to
illustrate the methodology for New Physics contributions and the 
relation of this to rare $D^0$ decays is pointed out. Finally, 
I address the subject of CP-violating asymmetries by describing 
some suggestions for future experimental studies and a recent 
theoretical analysis of New Physics contributions.
\end{abstract}

\maketitle

\thispagestyle{fancy}


\section{Introduction}
We are in the year following the momentous announcement of 
experimental evidence for $D^0$-${\bar D}^0$ 
mixing~\cite{{Aubert:2007wf},{Staric:2007dt},
{Abe:2007rd},{:2007uc}}.\footnote{Throughout the talk 
I will often refer to $D^0$-${\bar D}^0$ mixing as `charm mixing' 
or simply `$D^0$ mixing'.}  The time elapsed from the discovery 
of the charm degree of freedom was long (more than thirty years).  
From my own perspective, things really began to 
take off on the experimental side with the large fixed target 
experiments E687 and E791 in the 1990's.  Then the announcement 
by FOCUS of a much larger than expected width difference 
($y_{\rm CP} \sim 3.4\%$) created a buzz in the field.  Finally 
in 2007 we saw the positive results of the B-factory experiments 
BaBar and Belle (followed by CDF).   In the next Section, I 
review the latest findings and discuss what lessons they teach us.  
\section{Current Status of Charm Mixing}
Let us consider four points regarding the current experimental 
situation as summarized in Ref.~\cite{Schwartz:2008wa}.
\begin{enumerate}
\item {\it Experimental evidence for charm mixing has improved}.  
At the CHARM 2007 Workshop (Cornell University 8/5/07-8/8/07)
the euphoria of the moment provoked me to 
point out~\cite{Golowich:2007fs} that 
in light of the Physical Review Letters criteria of `observation' 
($> 5\sigma$) or `evidence' ($3\sigma$-to-$5\sigma$), the
then-existing $2.4\sigma$ determination of $x_{\rm D}$  
amounted to merely a `measurement' ($<
3\sigma$).
The current values (for the `no CP-violation' fit) 
\begin{eqnarray}
& & x_{\rm D} \equiv {\Delta M_{\rm D} \over \Gamma_{\rm D}} = 
0.98^{+0.26}_{-0.27}~\% 
  \nonumber \\
& & y_{\rm D} \equiv {\Delta \Gamma_{\rm D} \over 2 \Gamma_{\rm D}} =
\left( 0.75 \pm 0.18 \right)~\%  
\label{xy2008}
\end{eqnarray}
are at the level of `evidence', and indeed in a plot of 
$y_{\rm D}~vs~x_{\rm D}$,  the point 
$y_{\rm D} = x_{\rm D} = 0$ is excluded by 
$6.7\sigma$~\cite{Schwartz:2008wa}. 
\item {\it The current data set contains no evidence for 
CP-violation (hereafter CPV) in charm mixing}.  We will 
consider the corresponding situation for charm decays in Sect.~IV.  
\item {\it Weeding out theoretical descriptions}: 
Due to the heretofore uncertain status of charm mixing, 
I have been reluctant to discard various 
theoretical descriptions.   However, in view of the 
$95\%$ C.L. values in the CPV-allowed fit  to charm mixing,  
$0.39 \to 1.48$ for $x_{\rm D}(\%)$ and 
$0.41 \to 1.13$ for $y_{\rm D}(\%)$, I now feel that 
models having $y_{\rm D}, x_{\rm D} \sim 0.1\%$ are no longer 
tenable.  
\item {\it The strong phase $\delta$ is not `very large'}.  
The no-CPV determination yields 
$\delta(^{\rm o}) = 21.6^{+11.6}_{-12.6}$ whereas 
in the CPV-allowed fit the $95\%$ C.L. values are 
$-6.3 \to 44.6$ for $\delta(^{\rm o})$.  This developing 
topic is detailed in the following subsection.
\end{enumerate}
\subsection{The Strong Phase}
In the field of charm mixing, the `strong phase' is defined as the 
relative phase between the $D^0\to K^+\pi^-$ and $D^0\to K^-\pi^+$
decay amplitudes.  It appears in wrong-sign $D^0$ transitions 
because the $K^+\pi^-$ final state occurs both via doubly 
Cabibbo suppressed (DCS) decays and $D^0$ mixing followed by a 
Cabibbo favored (CF) decay.  As a consequence, the parameters 
\begin{eqnarray}
& & x_{\rm D}^\prime \equiv x_{\rm D} \cos\delta + y_{\rm D} \sin\delta
  \nonumber \\
& & y_{\rm D}^\prime \equiv y_{\rm D} \cos\delta - x_{\rm D} \sin\delta
\label{strph1}
\end{eqnarray}
appear in the analysis.  In the world of flavor SU(3) invariance, one 
has $\delta = 0$.  

There is no way to completely avoid the presence of $\delta$.  For
example, the time dependent rate for wrong-sign events in $D^0$ decay, 
\begin{eqnarray}
& & R\left({t\over\tau}\right) = R_{\rm D} + \sqrt{R_{\rm D}} 
y_{\rm D}^\prime {t \over \tau} + 
{x_{\rm D}^{\prime 2} +  y_{\rm D}^{\prime 2} \over 4} 
{t^2 \over \tau^2}\ ,
\label{strph2}
\end{eqnarray}
depends explicitly on $x_{\rm D}^\prime$ and $y_{\rm D}^\prime$.  
There is no fundamental physics in $\delta$; it is a detail of the 
strong interactions.  From the viewpoint of electroweak physics 
(and more specifically of charm mixing), it is an irritant, a nuisance.

Let me briefly review two rather different theoretical attempts to 
determine $\delta$.  Although neither is a first-principles
application of QCD, both are well thought-out calculations.
\begin{enumerate}
\item {\it Nearby Resonances}: In Ref.~\cite{Falk:1999ts},  
the CF and DCS amplitudes, respectively called $A$ and $B$, 
are each expressed as a sum of tree and resonance contributions.  
The latter involves the weak transition of the initial $D^0$ 
into a resonance $K^*$ whose mass is nearby that of the 
$D^0$~\cite{Golowich:1998pz}.  
The $K^*$ propagates and then decays strongly into the final state.
The relative phase between tree and resonance components arises 
from the phase $\phi$ of the $K^*$ propagator, 
\begin{eqnarray}
& & \tan\phi = - {\Gamma_{K^*} M_{\rm D} \over M_{\rm D}^2 - 
M_{\rm K}^{*2} } \ \ .
\label{strph3}
\end{eqnarray}
Straightforward algebra then relates the strong phase $\delta$ to 
the resonance phase $\phi$.  In view of the vanishing 
of $\delta$ in the SU(3) limit, it is convenient to plot $\sin\delta$ as 
a function of an SU(3)-breaking parameter $R_{\rm exp}$, 
\begin{eqnarray}
& & R_{\rm exp} \equiv {{\cal B}_{D^0 \to K^+\pi^-} \over 
{\cal B}_{{\bar D}^0 \to K^+\pi^-}} \cdot \left| 
{V_{\rm ud}~ V_{\rm cs} \over  V_{\rm us}~ V_{\rm cd}}\right|^2 \ \ .
\label{strph4}
\end{eqnarray}
At the time Ref.~\cite{Falk:1999ts} was written, one had 
$R_{\rm exp}^{\rm (1999)} = 1.58 \pm 0.49$.  This led to speculation 
among some that $\delta$ was quite large, although the large
uncertainty in $R_{\rm exp}$ allows no such conclusion.  
A more recent evaluation gives 
$R_{\rm exp}^{\rm (2008)} \simeq 1.2 \pm 0.04$.  The uncertainty is now 
rather smaller and so is the central value.  
\item {\it Phenomenological $D\to K\pi$ Analysis}: 
In Ref.~\cite{Gao:2006nb}, 
a study of seven CF and DCS $D \to K\pi$ modes is carried out 
in a model based on a traditional factorization and quark diagram 
approach.  SU(3) breaking is incorporated largely via the decay constants 
$f_K$ and $f_\pi$.  A formula for $\cos\delta$ is derived in terms 
of the branching ratios ${\cal B}_{{\bar D}^0 \to K^-\pi^+}$, 
${\cal B}_{{\bar D}^0 \to K^+\pi^-}$ and 
${\cal B}_{{\bar D}^+ \to K^+\pi^0}$. 
\end{enumerate}
From the above two models, we should not be suprised to find 
$|\delta| \le 20^o$ (both approaches predict only 
the magnitude $|\delta|$).  In other words, the strong phase 
is not expected to be `very large', a result in accord with the 
phenomenological analysis of Ref.~\cite{He:2007aj} (whose determinaton 
gives a result consistent with zero).

There has been recent progress in measuring $\delta$ experimentally. 
Consider the reaction chain~\cite{Asner:2005wf} 
\begin{eqnarray}
& & e^+e^- \to \psi(3770) \to D^0{\bar D}^0 \to ij\ ,
\label{strph5}
\end{eqnarray}
where $ij$ refers to some final state.  The $D^0{\bar D}^0$ pair 
will have ${\cal C} = {\cal P} = -1$.  Define the CP eigenstates 
\begin{eqnarray}
& & {\cal C} {\cal P} |D_1 \rangle = - |D_1 \rangle \ , \qquad 
{\cal C} {\cal P} |D_2 \rangle = + |D_2 \rangle \ \ .
\label{strph6}
\end{eqnarray}
The approach under discussion has the two remarkable features of 
emphasizing quantum correlations and of directly involving the 
CP-eigenstates.  The transition rate for producing the final state 
$ij$ obeys
\begin{eqnarray}
& & \Gamma(i,j) \propto \left|\langle i|D^0 \rangle \langle j | 
{\bar D}^0 \rangle \ - \ \langle j|D^0 \rangle \langle i | 
{\bar D}^0 \rangle \right|^2  \nonumber \\
& & \hspace{0.98cm} = \left|\langle i|D_2 \rangle \langle j | 
D_1 \rangle \ - \ \langle j|D_2 \rangle \langle i | 
D_1 \rangle \right|^2  \ \ .
\label{strph7}
\end{eqnarray}
The minus sign (since ${\cal C} = -1$) in the first of the rate 
equations is due to the quantum nature of the process and 
the second of the rate equations exhibits the explicit presence of 
the CP-eigenstates $D_{1,2}$.  Not all final states $ij$ are optimal 
for determining the strong phase.  It is best to choose one of the 
final states as a CP eigenstate $S_\pm$ and the other a $K\pi$ pair, 
{\it e.g.} as in
\begin{eqnarray}
& & F^{\rm corr}_{S_+/K\pi} = 
\left|\langle S_+|D_2 \rangle \langle K^-\pi^+|D_1\rangle\right|^2 
\nonumber \\
& & \hspace{1.27cm} = A_{S_+}^2 A_{K^-\pi^+}^2 
\left|1 + r e^{-i\delta}\right|^2 
\label{strph8}
\end{eqnarray}
where the dependence on $\delta$ is explicit.  Measurements based 
on this approach are presented in 
Refs.~\cite{{Asner:2008ft},{Rosner:2008fq}}, which give 
$\cos\delta = 1.03 \pm 0.31 \pm 0.06$.   By further including 
external measurements of charm mixing parameters, an alternate 
measurement of $\cos\delta$ is obtained, yielding 
$\delta = (22^{+11}_{-12})^o$.  Presumably, future 
experimental studies will be able to reduce present uncertainties, 
{\it e.g.} for a discussion of measuring the strong phase at 
BES-III see Ref.~\cite{Cheng:2007uj}.  

\section{The Origin of Charm Mixing}
In principle, charm mixing can arise from the Standard Model (SM) 
and/or from New Physics (NP).  We shall cover both in this talk, but 
consider the NP possibility in greater detail.

\subsection{Standard Model}
Theoretical estimates of charm mixing have been performed using 
either quark or hadron degrees of freedom.  We shall discuss each 
of these in turn.

\subsubsection{Quark Degrees of Freedom}


To my knowledge, the earliest attempt of this type is continaed in 
Ref.~\cite{datta}.  These days, the usual approach 
(like that used in $B_{d,s}$ mixing) is to express 
the mixing matrix element as a sum of local operators 
ordered according to dimension (operator product expansion or simply 
OPE)~\cite{Georgi:1992as}.  At a given order in the OPE, the mixing 
amplitude is expanded 
in QCD perturbation theory.  Finally matrix elements of the various 
local operators are determined.  It is a peculiarity of charm mixing 
that the various mixing amplitudes are most conveniently 
characterized by expanding 
in the small parameter $z \equiv (m_s/m_c)^2 \simeq 0.006$.  

A full implementation of this program is daunting because 
the number of local operators increases sharply with the 
operator dimension $D$ ({\it e.g.} $D=6$ has two operators, 
$D=9$ has fifteen, and so on)~\cite{Ohl:1992sr}.  
The matrix elements of the various local operators are unknown 
and can be only roughly approximated in model calculations.  
In principle, QCD lattice determinations would be of great use, 
but they are not generally available at this time~\cite{Gupta:1996yt}.  

An analysis of $x_{\rm D}$ and $y_{\rm D}$ in the leading order $D=6$ 
in the OPE has been carried out through ${\cal O}(\alpha_s)$ in 
Ref.~\cite{Golowich:2005pt}.  The result through ${\cal O}(\alpha_s)$ 
is $x_{\rm D} \simeq y_{\rm D} \sim 10^{-6}$.  These small values 
are due in part to severe flavor cancellations (the leading terms in 
the $z$-expansion for $x_{\rm D}$ and $y_{\rm D}$ respectively are 
$z^2$ and $z^3$ at order $\alpha_s^0$ and $z^2$ and $z^3$ 
at order $\alpha_s^1$.  

Evidently, the quark approach as implemented via the OPE has been 
seen as {\it not} the way to understand charm mixing.  It involves 
a triple expansion (in operator dimension $D$, QCD coupling 
$\alpha_s$ and parameter $z$) which is at best slowly convergent.  
One long-standing beacon of hope has been the suggestion in 
Ref.~\cite{Bigi:2000wn} that six-quark operators whose Wilson 
coefficients suffer only one power of $z$ suppression might  
give rise to $x_{\rm D} \sim 0.1\%$.  Even this effect 
(whose estimated size is problematic due to uncertainties in matrix 
element evaluation) is too small.  

\subsubsection{Hadronic Degrees of Freedom}


Let us restrict our attention to the following exact relation for 
the width difference $\Delta\Gamma_{\rm D} = {\rm Im}\,I/M_{\rm D}$ 
where 
\begin{eqnarray}
& & I \equiv \langle {\bar D}^0 |
    \,i\! \int\! {\rm d}^4 x\, T \Big\{
    {\cal H}^{|\Delta C|=1}_w (x)\, {\cal H}^{|\Delta C|=1}_w(0) \Big\}
    | D^0 \rangle \ . \nonumber \\
\label{had1}
\end{eqnarray}
We can calculate $y_{\rm D}$ by inserting intermediate states between the 
$|\Delta C| = 1$ weak hamiltonian densities ${\cal H}^{|\Delta C|=1}_w$.  
Of course, knowledge of the matrix elements $\langle n |  
{\cal H}^{|\Delta C|=1}_w | D^0 \rangle$ is required.  
This method yielded an entirely reasonable estimate for 
$y_{{\rm B}_s}$, where the number of large matrix elements turns out
to be quite limited~\cite{Aleksan:1993qp}.  

By contrast, for charm mixing the number of contributing matrix 
elements is quite large.Perhaps the most comprehensive analysis to 
date for charm is the phenomenological evaluation based on
factorization given in Ref.~\cite{Buccella:1994nf}.  The result 
$y_{\rm D} \sim 0.1\%$ thus obtained is too small.  

This unfortunate circumstance shows how delicate this sum over many 
contributions seems to be.  What then is one to do, given that 
the hadron approach appears tied to the issue of matrix element 
evaluation?  Perhaps it is best to rely more on charm decay data 
and less on the underlying theory.  The earliest work of this 
type~\cite{Donoghue:1985hh,lw} focussed on the $P^+P^- = \pi^+\pi^-, 
K^+ K^-, K^- \pi^+, K^+ \pi^-$ intermediate states in 
Eq.~(\ref{had1}).  
In the flavor SU(3) limit, exact cancellations reduce the 
contribution from this subset of states to zero.  However, SU(3) 
breaking was already known to be significant in individual 
charm decays, and based on data available at that time, 
these references concluded that `$y_{\rm D}$ might be large'.  

A modern version of this approach exists~\cite{Falk:2004wg}, but 
with the above argument essentially turned on its head.  A main 
new ingredient is the realization that SU(3) breaking occurs at 
second order in charm mixing~\cite{Falk:2001hx}.  Can it be that 
all two-particle and 
three-particle sectors (such as $P^+P^-$) contribute very little 
to charm mixing due to flavor cancellations?  Perhaps, but this 
argument cannot be continued to the four-particle intermediate states 
because decay into four-kaon states is kinematically forbidden.  
In fact, Ref.~\cite{Falk:2004wg} claims that these very sectors  
can generate $y_D\sim 10^{-2}$.  A dispersion relation calculation 
using charm decay widths as input can be used to estimate $x_{\rm D}$, 
but this contains an additional layer of model dependence. 

I believe the claim that SM contributions produce 
values for $x_{\rm D}$ and $y_{\rm D}$ at the $1\%$ level is not 
unreasonable.  At the same time, however, compared to SM predictions 
for kaon, $B_d$ and $B_s$ mixing, the status of charm mixing 
is decidely `fuzzy'. 

\subsection{New Physics}

\begin{figure}[b]
\centering
\includegraphics[width=40mm]{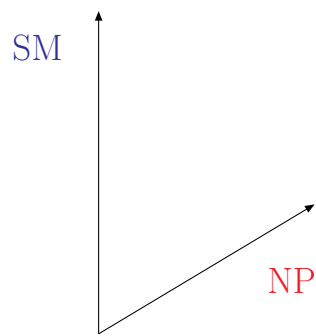}
\caption{Possibility of SM and NP Contributions.} \label{fig:smnp}
\end{figure}

The LHC era is about to begin.  Yet, what we will learn from LHC data 
is still highly uncertain.  This is in stark contrast with SM 
expectations at the time LEP came on line.  Our own recent work on 
$x_{\rm D}$ has tried to be bias-free by allowing for a variety of 
extensions to the Standard 
Model~\cite{Golowich:2007ka},

\vspace{0.3cm} 

1] Extra gauge bosons (LR models, {\it etc})

2] Extra scalars (multi-Higgs models, {\it etc})

3] Extra fermions (little Higgs models, {\it etc})

4] Extra dimensions (split fermion models, {\it etc})

5] Extra global symmetries (SUSY, {\it etc}).

\vspace{0.2cm} 

NP contributions to charm mixing can affect $y_{\rm D}$ as well as 
$x_{\rm D}$.  We do not consider 
the former in this talk, but instead refer the reader to 
Refs.~\cite{{Golowich:2006gq},{Petrov:2007gp}}.

The strategy for calculating the effect of NP on $D^0$ mixing 
is, for the most part, straightforward.  One considers a 
particular NP model and calculates the mixing amplitude for 
as a function of the model parameters.  If the mixing signal 
is sufficiently large, constraints on the parameters are obtained.  
For all we know, the observed $D^0$ mixing signal is a product 
of both SM {\it and} NP contributions.  In general we will not 
know the relative phase between the SM and NP amplitudes, as 
depicted in Fig.~\ref{fig:smnp}, or even the precise value of 
the `fuzzy' SM component.  This affects how NP 
constraints are treated, as shown later in a specific example.

We now turn to the issue of NP and $x_{\rm D}$, as based on 
the work in Ref.~\cite{Golowich:2007ka}, which studied a total 
of 21 NP models.  These are listed in Table~\ref{tab:bigtable}.

\begin{table}[b]
\caption{NP models studied in Ref.~\cite{Golowich:2007ka}}
\begin{tabular}{|c|}
\colrule\hline 
Model 
\\ \hline\hline
Fourth Generation  \ \ \\
$Q=-1/3$ Singlet Quark  \\
$Q=+2/3$ Singlet Quark  \\
Little Higgs  \\ 
Generic $Z'$ \\
Family Symmetries \\
Left-Right Symmetric  \\ 
Alternate Left-Right Symmetric \\ 
Vector Leptoquark Bosons \\
\ \ Flavor Conserving Two-Higgs-Doublet \ \ \\
Flavor Changing Neutral Higgs  \\
\ \  FC Neutral Higgs (Cheng-Sher ansatz) \ \  \\
Scalar Leptoquark Bosons  \\
Higgsless \\
Universal Extra Dimensions \\
Split Fermion  \\
Warped Geometries \\
Minimal Supersymmetric Standard \\
Supersymmetric Alignment \\
Supersymmetry with RPV \\
Split Supersymmetry \\ 
\hline\hline
\end{tabular}
\vskip .05in\noindent
\label{tab:bigtable}
\end{table}

Of these 21 NP models, only four (split SUSY, universal extra
dimensions, left-right symmetric and flavor-changing two-higgs 
doublet) are ineffective in producing charm mixing at the observed 
level.  This has several causes, {\it e.g.} the NP mass scale is too 
large, severe cancellations occur in the mixing signal, {\it etc}.  
This means that 17 of the NP models {\it can} produce charm mixing.  
For these, we can get constraints on masses and mixing parameters. 

\subsubsection{R$_{\rm P}$ Violations and $D^0$ Mixing}

\begin{figure}[h]
\centering
\includegraphics[width=80mm]{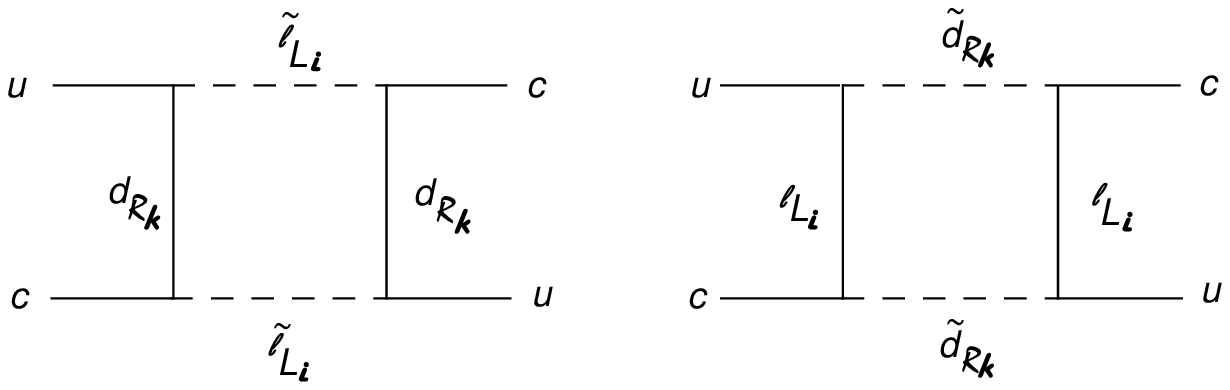}
\caption{RPV box diagram.} \label{fig:rpvbox}
\end{figure}

We cannot review all 17 NP models here, so we shall conentrate on 
just one of them, the case of R-parity violating (RPV) supersymmetry.  
RPV contributes to $D^0$ mixing via box amplitudes, as displayed in 
Fig.~\ref{fig:rpvbox}.  Each box diagram is seen to contain four 
vertices in which quarks, squarks and leptons interact.  

\begin{figure}[b]
\centering
\includegraphics[width=82mm]{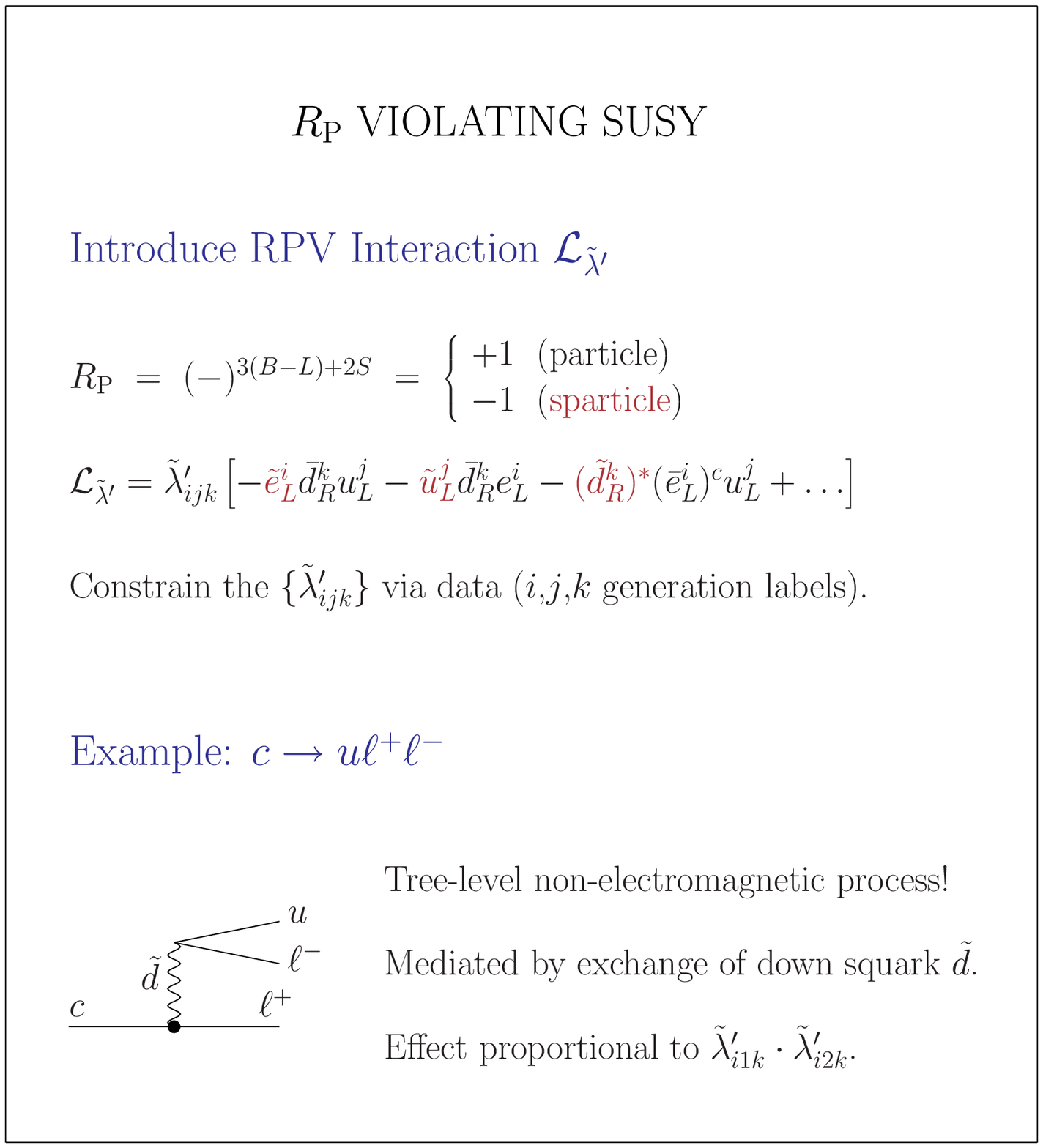}
\caption{Outline of R-parity violation.} \label{fig:rpv1}
\end{figure}

Fig.~\ref{fig:rpv1} provides a brief summary about this topic.
The quantum number R$_{\rm P}$ distinguishes between particles of the SM 
and their supersymmetric partners (`sparticles').  R-parity need not 
be conserved and in Fig.~\ref{fig:rpv1} we display an 
R$_{\rm P}$-violating lagrangian that is relevant to charm-mixing.  
The coupling strength is ${\tilde \lambda}^\prime_{ijk}$ and the 
indices $i,j,k = 1,2,3$ are generation labels.  The amplitude for 
$c\to u \ell^+\ell^-$ appearing in Fig.~\ref{fig:rpv1} is part of 
the box-diagram which mediates the charm mixing.  Note that it 
is proportional to the product 
${\tilde \lambda}^\prime_{i1k}~{\tilde \lambda}^\prime_{i2k}$.  
Incidentally, to an experimentalist who has dealt with 
lepton-antilepton pairs, this amplitude has the unexpected feature 
that the pair come from distinct vertices and not from a photon
or a Z$^0$.  

\begin{figure}[t]
\centering
\includegraphics[width=50mm,angle=90]{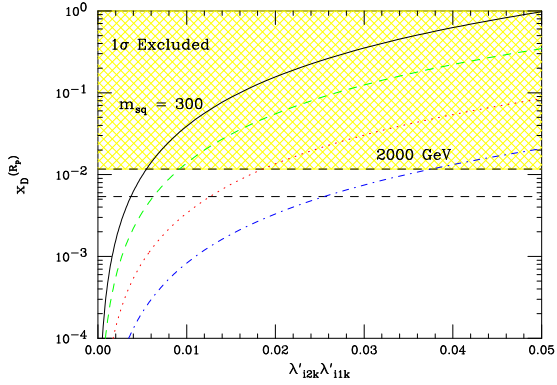}
\caption{Constraints on RPV from $D^0$ mixing.} \label{fig:rpv4}
\end{figure}

We refer the reader to Ref.~\cite{Golowich:2007ka} for details 
regarding the calculation of the RPV contributions to $D^0$ mixing.  
The end result of the analysis is the set of constraints 
displayed in Fig.~\ref{fig:rpv4}.  There, $x_{\rm D}$ is plotted 
as a function of the product of the RPV couplings 
$\tilde\lambda'_{i2k}\tilde\lambda'_{i1k}$.  We take 
$m_{\tilde d_{R,k}}=\
m_{\tilde\ell_{L,i}}$, with $m_{\tilde d_{R,k}}=300\,, 500\,, 1000$
and 2000 GeV corresponding to the solid, green dashed, red dotted, and
blue dashed-dot curves, respectively.  The $1\sigma$ experimental 
bounds are as indicated, with the yellow shaded region depicting the
region that is excluded.  The bound cited in
Ref.~\cite{Golowich:2007ka} for the RPV couplings is 
\begin{eqnarray}
& & \tilde\lambda'_{i2k}\tilde\lambda'_{i1k}\leq 0.085 
\sqrt{x_{\rm D}^{\rm (expt)}} 
\left({m_{\tilde d_{R,k}}}\over 500~{\rm GeV}\right)
\label{bnd1}
\end{eqnarray} 
or using the updated mixing value of Eq.~(\ref{xy2008}) we find 
\begin{eqnarray}
\tilde\lambda'_{i2k}\tilde\lambda'_{i1k} 
\leq 1.7 \cdot 10^{-3} m_{\tilde d_{R,k}}/100~{\rm GeV}\ \ .
\label{bnd2}
\end{eqnarray}

\subsubsection{The Rare Decay $D^0 \to \mu^+\mu^-$} 
It is often profitable for phenomenological studies to encompass 
various physical processes at the same time.  The following is a 
case in point.  It should be clear that the squark-exchange diagram 
at the bottom of Fig.~\ref{fig:rpv1} contributes not only to 
vertices in the $D^0$ mixing amplitude but also to rare transitions 
such as $D^0 \to \mu^+\mu^-$, $D^0 \to \pi^+\mu^+\mu^-$, 
{\it etc}~\cite{Burdman:2001tf}.  In fact,  the current experimental 
limit~\cite{pdg}  
\begin{eqnarray*}
{\cal B}_{D^0 \to \mu^+\mu^-} \le 1.3 \cdot 10^{-6} 
\end{eqnarray*}
implies the constraint 
\begin{eqnarray}
& & \tilde\lambda'_{21k}\tilde\lambda'_{22k}\leq
7 \cdot 10^{-3} m_{\tilde d_{R,k}}/100~{\rm GeV} \ \ .
\label{bnd3}
\end{eqnarray}
This ilustrates 
how charm mixing and charm rare decays are both of interest to RPV  
phenomenology.  They are roughly competitive at present.  However, 
the limit from charm isnot going to change much whereas that from 
$D^0 \to \mu^+\mu^-$ can continue to improve.

\section{CPV Asymmetries}
Since there is no existing evidence for CPV in the charm sector, it is 
natural to look to the future.  We consider two topics of this type, 
first, possible experimental strategies for detecting CPV signals and 
next, a survey of NP models and their CPV asymmetries.   For existing 
literature on the subject, I recommend a recent discussion by 
Petrov~\cite{Petrov:2007ms} and a treatment of basic CPV formalism 
applied to charm by Xing~\cite{Xing:1996pn}. 

\subsection{Future Strategies}
I briefly review two papers, each involving a facility 
planned for future operation. 
\begin{enumerate}
\item {\it Super B-factory}: 
A suggestion for work at a super B-factory is to probe 
charm mixing and CPV using coherent $D^0 {\bar D}^0$ events from 
$\Upsilon (1S)$ decays~\cite{Li:2006wx}.  The point is that the 
large boost factor
in the $\Upsilon (1S)$ rest frame ($\simeq 2.33$) would allow a
precise determination ofthe proper time interval $\tau$ between the
two $D$ decays.  Thus for a final state $f_1f_2$, one would measure 
\begin{eqnarray}
& & R(f_1,f_2;t) = {d\Gamma_{\Upsilon_{1S} \to f_1f_2} 
\over d\tau} \ \ .
\end{eqnarray}
Symmetric and asymmetric $\Upsilon (1S)$ factories
are considered and various CPV asymmetries discussed.  A yield of 
$10^7 \to 10^8$ D-pairs per year is estimated.
\item {\it $\tau$-Charm Factory}: 
Ref.~\cite{Xing:2007sd} considers the decay modes $D \to K^* K$ obtained by
running on the $\psi(3770)$ and $\psi(4140)$ resonances at a 
$\tau$-Charm factory.  Note that the C-parity values differ for these 
two states, with ${\cal C}[\psi(3770)] = -1$ and ${\cal C}[\psi(3770)]
= +1$.  The production of final states for such experiments would be 
coherent, and one defines the quantities 
\begin{eqnarray}
& & \Gamma^{++}_{\cal C} \equiv \Gamma (K^{*+}K^-, K^{*+}K^-)_{\cal C}
\nonumber \\
& & \Gamma^{--}_{\cal C} \equiv \Gamma (K^{*-}K^+, K^{*-}K^+)_{\cal
  C} 
\nonumber \\
& & \Gamma^{+-}_{\cal C} \equiv \Gamma (K^{*+}K^-, K^{*-}K^+)_{\cal C}
\nonumber \\
& & \Gamma^{-+}_{\cal C} \equiv \Gamma (K^{*-}K^+, K^{*+}K^-)_{\cal
  C} \ \ .
\end{eqnarray}
The authors conclude that it is favorable to measure decays of 
correlated $D$'s the various $K^* K$ states by running on the
$\psi(4140)$, with a candidate CPV observable being $(\Gamma^{++}_+ 
- \Gamma^{--}_+)/\Gamma^{+-}_+$.
\end{enumerate}

\subsection{Calculating CPV Asymmetries}
An interesting analysis of CP-violations in the 
singly-Cabibbo-suppressed transitions 
\begin{eqnarray*}
& & c \ \to \ u~{\bar q}~q \qquad (q = s,d) 
\end{eqnarray*}
as recently been carried out in 
Ref.~\cite{Grossman:2006jg}.  
Final states which are both CP eigenstates ($K^+K^-,\pi^+\pi^-,
{\it etc}$) and non-CP eigenstates ($K^*K,\rho\pi, {\it etc}$)
are considered.  

Let us restrict our attention to the time-integrated 
CPV asymmetries of a final state $f$ which is a CP eigenstate 
(${\bar f} = f$),
\begin{eqnarray}
& & a_f \equiv {\Gamma_{D^0 \to f} - \Gamma_{{\bar D}^0 \to f} 
\over \Gamma_{D^0 \to f} + \Gamma_{{\bar D}^0 \to f} } \ \ .
\end{eqnarray}
Such an asymmetry can receive contributions from decay, mixing 
and interference, 
\begin{eqnarray}
& & a_f \ = \ a_f^d + a_f^m + a_f^i \ \ .
\end{eqnarray}
The `direct' component ({\it i.e.} from decay) is generally expressed
as 
\begin{eqnarray}
& & a_f^d \ = \ 2 r_f \sin\phi_f \sin\delta_f \ \ , 
\end{eqnarray}
where the phases $\phi_f$ and $\sin\delta_f$ arise respectively 
from CPV and QCD.  

What are the experimental prospects for measuring such 
CPV asymmetries?  There can, in principle, be both SM and 
NP components.  As with charm mixing, 
there can be both short-distance and long-distance SM 
contributions, with the latter subject to less suppression 
than the former.  However, due to uncertainties in estimating 
the long-distance component it is hard to be very precise about 
the actual size of SM asymmetries.  At any rate, 
it is concluded in Ref.~\cite{Grossman:2006jg} 
that the SM cannot generate CPV asymmetries in the SCS sector 
much larger than ${\cal O}(10^{-4})$. 

At the time at which the work of Ref.~\cite{Grossman:2006jg} was 
carried out, the scale of experimental limits on the CPV 
asymmetries was roughly ${\cal O}(10^{-2})$.  This would appear 
to present a wide window of opportunity for observation of NP effects.  
However, some up-to-date (as of 1/31/08) experimental limits from the 
Charm Heavy Flavor Averaging Group~\cite{hfag} are exhibited in 
Table~\ref{tab:asymm}.  In arranging this Table, I selected only 
those limits whose uncertainties are less than $0.01$.  We see 
that the window is not as large today!  

\begin{table}[h]
\begin{center}
\caption{Current CPV Asymmetry Data}
\begin{tabular}{|c|c|c|}
\hline \textbf{Asymmetry} & \textbf{Mode} & \textbf{Value} 
\\
\hline  
 & $D^0 \to K^+K^-$ & $0.0015 \pm 0.0034$\\
 & $D^0 \to \pi^+\pi^-$ & $0.0002 \pm 0.0051$\\
$\Delta\Gamma/2\Gamma$ & $D^0 \to K^- \pi^+\pi^0$ 
& $0.0016 \pm 0.0089$\\
 & $D^+ \to K_{\rm S}\pi^+$ & $0.0086 \pm 0.009$  \\
 & $D^+ \to K^+K^-\pi^+$  & $0.0059 \pm 0.0075$  \\
\hline 
$\Delta\tau/2\tau$ & $D^0 \to K^+K^-,~\pi^+\pi^-$ 
& $0.0012 \pm 0.0025$ \\
\hline
\end{tabular}
\label{tab:asymm}
\end{center}
\end{table}

I leave it to the reader to study the various technical details present 
in Ref.~\cite{Grossman:2006jg}.  However, some general patterns 
of expected NP behavior are:
\begin{enumerate}
\item Some supersymmetric models can give $a_f^d \sim 
{\cal O}(10^{-2})$ whereas models with minimal flavor violation 
cannot.
\item Only the SCS decays probe gluonic penguin amplitudes.  
Thus any large CPV asymmetries arising from this source would 
be unlikely for CF and DCS decays.
\item CPV asymmetries as large as $a_f^d \sim {\cal O}(10^{-2})$ 
would be expected from NP theories which contribute via loop 
amplitudes but not tree amplitudes (tree amplitudes tend to be 
constrained by $D^0$ mixing constraints).
\end{enumerate}

%

\section{Conclusions}
As we enter the LHC era, our field will require the 
resources to pursue both {\it discovery} and {\it precision} options.  
The discovery option will be carried out at the LHC.  If, as
anticipated, New Physics is revealed, perhaps (i) the signature will 
be so striking that a specific NP model is clearly identified, or (ii) 
the situation will be unclear for quite some time ({\it e.g.} some of 
the NP degrees of freedom might remain beyond the LHC reach).  In
either case, it will be important to carry out the precision option.
For the case (i) above, we need to check and verify the LHC results, 
whereas for case (ii) observing the pattern of rare effects should 
help clarify the LHC findings.  This will require the participation 
of LHC-B and $e^+e^-$ super-flavor factories.  How many such
facilities will become operable only time will tell.  One can only
hope.  Now onto a summary of the main topics:

\vspace{0.3cm} {\it Charm mixing and experiment}: 

The data on $D^0$ mixing allow us at long last to claim (in 
the sense of PRL discovery criteria) 
`evidence' for determinations of $x_{\rm D}\ \&\  y_{\rm D}$ 
and a true `observation' of mixing.  The quality of the mixing 
signal now can rule out theoretical descriptions predicting 
charm mixing at the $0.1\%$ level.  There has been real progress on 
the issue of the strong phase difference $\delta$ between the 
$D^0\to K^-\pi^+$ and $D^0\to K^+\pi^-$ amplitudes.  We expect that 
improved sensitivity in the quantum correlation approach will 
provide a more accurate measure of $\delta$.  

\vspace{0.3cm} {\it Charm mixing and Standard Model theory}: 

There is little change in our previous understanding of this 
subject.  The quark approach which is carried out in the OPE 
has, to date, yielded $x_{\rm D}, y_{\rm D} \sim 10^{-6}$.  
Even the most optimistic prediction for using this method 
predicts a mixing signal an order of magnitude too small.  
We have here a very slowly convergent process which nobody 
has yet been able to conquer.  More promising is the hadron 
approach which might in fact get the magnitudes 
$x_{\rm D}, y_{\rm D} \sim 10^{-2}$ right, but is hampered by 
theoretical uncertainties.  Nonetheless, we are still able to 
conclude that the observed $D^0$ mixing might well be a 
consequence of SM physics.  

\vspace{0.3cm} {\it Charm mixing and New Physics theory}: 

The comprehensive study in Ref.~\cite{Golowich:2007ka} 
of 21 possible NP contributions to charm mixing shows 
that the observed $D^0$ mixing might also well be a 
consequence of beyond-SM physics!  Further progress on 
this front will presumably require input from LHC data 
for selecting among NP possibilities.  We have pointed out 
how the inter-related phenomenologies of charm mixing and rare 
charm decays allows for a more systematic probe of NP parameter spaces.

\vspace{0.3cm} {\it Studies involving CP-violations in charm}: 

With the observation of charm mixing, the study of CP-violations 
in charm has taken its place at the forefront of research in this 
field.  Given the expectation that CP-violating SM asymmetries 
should be less than ${\cal O}(10^{-3})$ and that some NP models can 
exceed this value, there should be a real window of opportunity 
to aim at.  However, this window has begun to close.  One is left 
wondering as usual -- where is the New Physics?



\begin{acknowledgments}
This work was supported in part by the U.S.\ National Science
Foundation under Grant PHY--0555304.  I wish to thank the 
Taiwan particle physics community for their fine job in organizing 
and carrying out FPCP 2008.
\end{acknowledgments}

\bigskip 

\end{document}